\documentclass[doublecol]{epl2} 

\newcommand{\be}{\begin{equation}}
\newcommand{\ee}{\end{equation}}
\newcommand{\bee}{\begin{eqnarray}}
\newcommand{\eee}{\end{eqnarray}}

\usepackage{amsmath,amssymb} 
\usepackage{dcolumn}
\usepackage{MnSymbol}	
\usepackage[capitalize]{cleveref}

\title{Shear-induced diffusion in non-local granular flows}
\shorttitle{Diffusion in non-local granular flows} 

\author{Prashidha Kharel\inst{1*} \and Pierre Rognon\inst{1**}}
\shortauthor{P. Kharel \and P. Rognon}

\institute{                    
	\inst{1} Particles and Grains Laboratory, School of Civil Engineering, The University of Sydney, Sydney, 2006 NSW, Australia\\
	\inst{*} Email: prashidha.kharel@sydney.edu.au\\
	\inst{**} Email: pierre.rognon@sydney.edu.au
}
\pacs{nn.mm.xx}{First pacs description}
\pacs{nn.mm.xx}{Second pacs description}
\pacs{nn.mm.xx}{Third pacs description}

\abstract{
We investigate the properties of self-diffusion in heterogeneous dense granular flows involving a gradient of stress and inertial number. 
The study is based on simulated plane shear with gravity and Poiseuille flows, in which non-local effects induce some creep flow in zones where stresses are below the yield. Results show that shear-induced diffusion is qualitatively different in zones above and below the yield. Below the yield, diffusivity is no longer governed by velocity fluctuations, and we evidenced a direct scaling between diffusivity and local shear rate. This is interpreted by analysing the grain trajectories, which exhibit a caging dynamics developing in zones below the yield. We finally introduce an explicit scaling for the profile of local inertial number in these zones, which leads to a straightforward expression of the diffusivity as a function of the stress and position in non-local flows. 
}

\begin{document}

\maketitle

\section{Introduction} 

Shearing dense granular flows induces diffusion of grains. This mechanism of shear-induced diffusion underpins the rate of mixing \cite{hsiau93}, heat transfer \cite{rognon_thermal_2010} and segregation \cite{ottino2000mixing} in a variety of natural and industrial granular flows. It is usually modelled by a coefficient of self-diffusion, also called diffusivity $D$ [m$^2$/s].

In homogeneous shear flows, in which there is no gradient of shear rate, three relationships have been established from which diffusivity can be predicted:

\bee
D &\propto& \delta v d \label{eq:Dscaling}\\
\frac{\delta v}{\dot \gamma d} &\propto& I^{-\frac{1}{2}}; \;\;\;I = \dot \gamma t_i \;\;\;t_i = d \sqrt{\rho/\sigma} \label{eq:Vscaling}\\
bI &=& \mu-\mu_s \text{ for } \mu > \mu_s; \;\;\; I = 0 \text{ otherwise.}  \label{eq:local}
\eee

The first scaling relates the diffusivity to the velocity fluctuations $\delta v$ and grain size $d$ \cite{zik1991self,natarajan1995local, hsiau2008mixing}. It is consistent with a typical grain trajectory following a random walk of step $d$ and frequency $\delta v/d$. The second scaling relates the velocity fluctuations to the inertial number $I$, itself comparing the shear rate $\dot \gamma$ to an inertial time $t_i$ involving the normal stress $\sigma$ in the flow, and the grain size $d$ and density $\rho$. It is consistent with the development of clusters of jammed grains of size $\ell/d \propto I^{-\frac{1}{2}}$ \cite{da2005rheophysics, DeGiuli:2015aa,PhysRevLett.119.178001}. At relatively high inertial numbers ($I\gtrsim 0.01$), this length scale reaches a minimum of $\ell =d$ and the velocity fluctuations are given by $\delta v \propto d \dot \gamma$. Accordingly, the diffusivity can be expressed as:

\be \label{eq:Dabove}
D \propto 
\begin{cases}
   d^2 \dot \gamma, & \text{for } I \gtrsim 0.01\\
     d^2 \dot \gamma \frac{1}{\sqrt{I}}, & \text{otherwise}
\end{cases}
\ee

The scaling (\ref{eq:local}) is a local constitutive law that relates the inertial number to the level of stresses within the flow. $\mu$ is the ratio of shear to normal stress, $\mu_s$ a yield criteria and $b$ a numerical constant \cite{da2005rheophysics,midi2004dense,Jop:2006aa}. Like Bingham fluids, this law indicates that there should be no flow ($\dot \gamma = 0$) if the shear stress is lower than a threshold, $\tau< \mu_s \sigma$. According to (\ref{eq:Dscaling}) and (\ref{eq:Vscaling}), there should then be no diffusion.

However, most granular flows are not homogeneously sheared, owing to some gradient of stresses. Then, non-local effects arise that cannot be captured by the local constitutive law alone. For instance, a flowing layer can induce some flow in a nearby layer where the stresses are below the yield ($\mu< \mu_0$) \cite{ jop_rheological_2015,mueth_signatures_2000,komatsu2001creep}. We refer to such layers as \textit{sub-yield} layers. 
A number of non-local models have been introduced to capture the profiles of inertial number in heterogeneous granular flows, including in sub-yield layers \cite{goyon_spatial_2008,bocquet2009kinetic,pouliquen_non_local_2009,kamrin2012nonlocal,bouzid2013nonlocal,wandersman2014nonlocal,miller2013eddy,rognon2015long}. In contrast, little is known about the shear-induced diffusion in these layers. Specifically, there is no evidence confirming or challenging the validity of the scalings (\ref{eq:Dscaling}) and (\ref{eq:Vscaling}) in sub-yield layers. 

The purpose of this Letter is to assess the validity of the diffusivity and velocity fluctuation scalings (\ref{eq:Dscaling}) and (\ref{eq:Vscaling}) in heterogeneous granular flows, with a special focus on sub-yield layers. In this aim, we have simulated a series of steady and heterogeneous granular flows in which non-local effects arise, and measured velocity fluctuations and diffusivity in different parts of the flows. 

\begin{figure}
    \centering
	\includegraphics[width=\columnwidth]{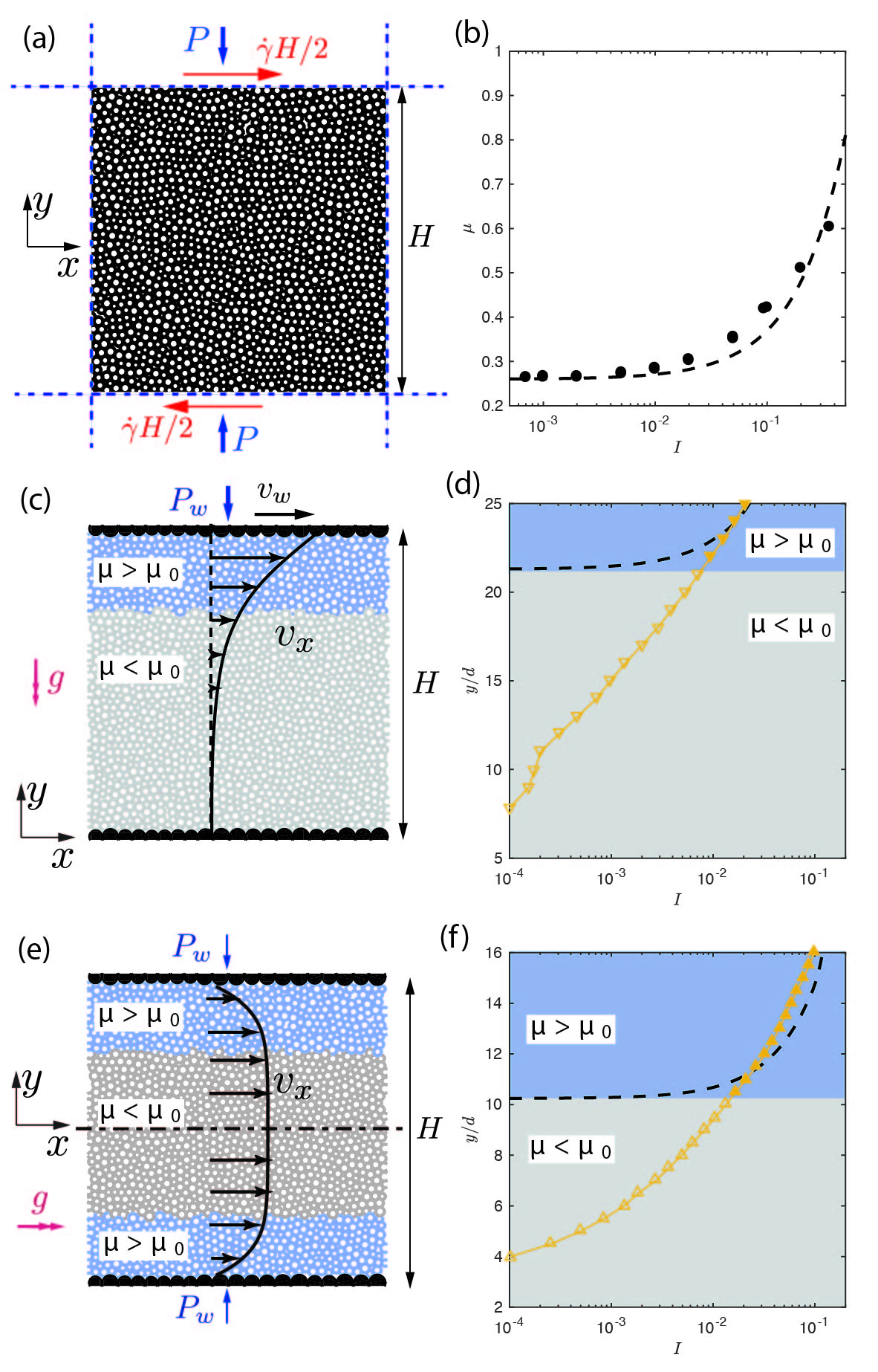}
    \centering
    \caption{Granular flows in different geometries. (a) schematic of plane shear without gravity (PS). (b) Local constitutive law: (symbols) shear stress ratio measured in plane shear flows at different prescribed inertial numbers;  (dashed lines) best fit using eq. (\ref{eq:local}) ($\mu_0=0.26$ and $b=1.1$).  (c) Schematic of plane shear with gravity (PSG) including a velocity profile. (d) Profile of inertial number measured in the PSG (symbols, see \cref{tab:config}) and prediction of the local constitutive law (dashed line). (e) Schematic of Poiseuille Flows (PF) with imposed lateral pressure, including a velocity profile. (f) Profile of inertial number measured in a PF (symbols) and the prediction of local constitutive law. In (c-f), blue regions are above the yield condition while grey regions are below the yield condition. In (d,f), the parameters of the flows are given in table \ref{tab:config}.} 
    \label{fig:sim_geo}
\end{figure}

\section{Simulated flows}

We use a Discrete Element Method to simulate dense granular flows in three geometries: plane shear (PS), plane shear with gravity (PSG), and Poiseuille flows (PF). 
These geometries are illustrated on figure \ref{fig:sim_geo} and detailed below.

All tests involve grains that are 2D disks of average diameter $d$, mass $m$ and density $\rho$. They interact by elastic, frictional and dissipative contacts characterised by a Young's modulus $E$, a coefficient of friction $\mu_g=0.5$ and a coefficient of restitution $e=0.5$ for normal impact. A uniform polydispersity of $\pm 20\%$ is introduced on the grain diameter to prevent crystallisation during shear. There is no contact adhesion and no fluid in the pore space. The value of these contact parameters only marginally influence the flow properties, as discussed in \cite{da2005rheophysics,DeGiuli:2015aa}.

\begin{table}[]
  \small
  \centering
    \caption{parameters of the simulated flows: Plane Shear (PS) from \cite{PhysRevLett.119.178001}, the Plane Shear with Gravity (PSG) and Poiseuille flow (PF), and corresponding symbols used in \cref{fig:sim_geo,fig:diffusion,fig:scaling,fig:prediction}. Filled symbols correspond to layers above the yield ($\mu(y)>\mu_0$) and open symbols correspond to sub-yield layers  ($\mu(y)<\mu_0$).}

    \begin{tabular}{c|c|c|c|c|c} 
         \label{tab:config}
      &&&&&\\
      Symbol & Geometry & $H/d $&$10^3\frac{P_w}{E}$ & $v_w \sqrt{\frac{\rho}{P_w}}$ & $\frac{g}{d}\sqrt{\frac{\rho}{P_w}}$\\
      &&&&&\\
      \hline
      \rlap{+}{\texttimes} & PS  &120&1 & - & 0 \\ 
      {\color[rgb]{0, 0.4470, 0.7410}$\blacksquare$},{\color[rgb]{0, 0.4470, 0.7410}$\square$} & PSG  &60&0.4 & 0.316 & 0.0095\\ 
      {\color[rgb]{0.8500,    0.3250,    0.0980}$\triangleleft$},{\large\color[rgb]{0.8500,    0.3250,    0.0980}$\blacktriangleleft$}  & PSG  & 60 & 0.4 & 0.791 & 0.0126\\ 
      {\color[rgb]{0.9290,    0.6940,    0.1250}$\blacktriangle$},{\color[rgb]{0.9290,    0.6940,    0.1250}$\triangle$} & PSG  &30 & 0.4 & 0.316 & 0.019\\ 
      {\color[rgb]{0.4940,    0.1840,    0.5560}$\triangleright$},{\large\color[rgb]{0.4940,    0.1840,    0.5560}$\blacktriangleright$} & PSG  &30 & 0.4 & 0.791 & 0.019\\
      {\color[rgb]{0, 0.4470, 0.7410}$\bullet$},{\color[rgb]{0, 0.4470, 0.7410}$\circ$} & PF &80& 1 & - & 0.01\\ 
      {\color[rgb]{0.8500,    0.3250,    0.0980}$\blacklozenge$},{\color[rgb]{0.8500,    0.3250,    0.0980}$\lozenge$} & PF & 80 & 1 & - & 0.0125\\ 
      {\color[rgb]{0.9290,    0.6940,    0.1250} $\blacktriangledown$},{\color[rgb]{0.9290,    0.6940,    0.1250} $\triangledown$} & PF &40 & 1 & - & 0.036\\ 
    \end{tabular}
\end{table}

\begin{figure}[]
    \centering
	\includegraphics[width=\columnwidth]{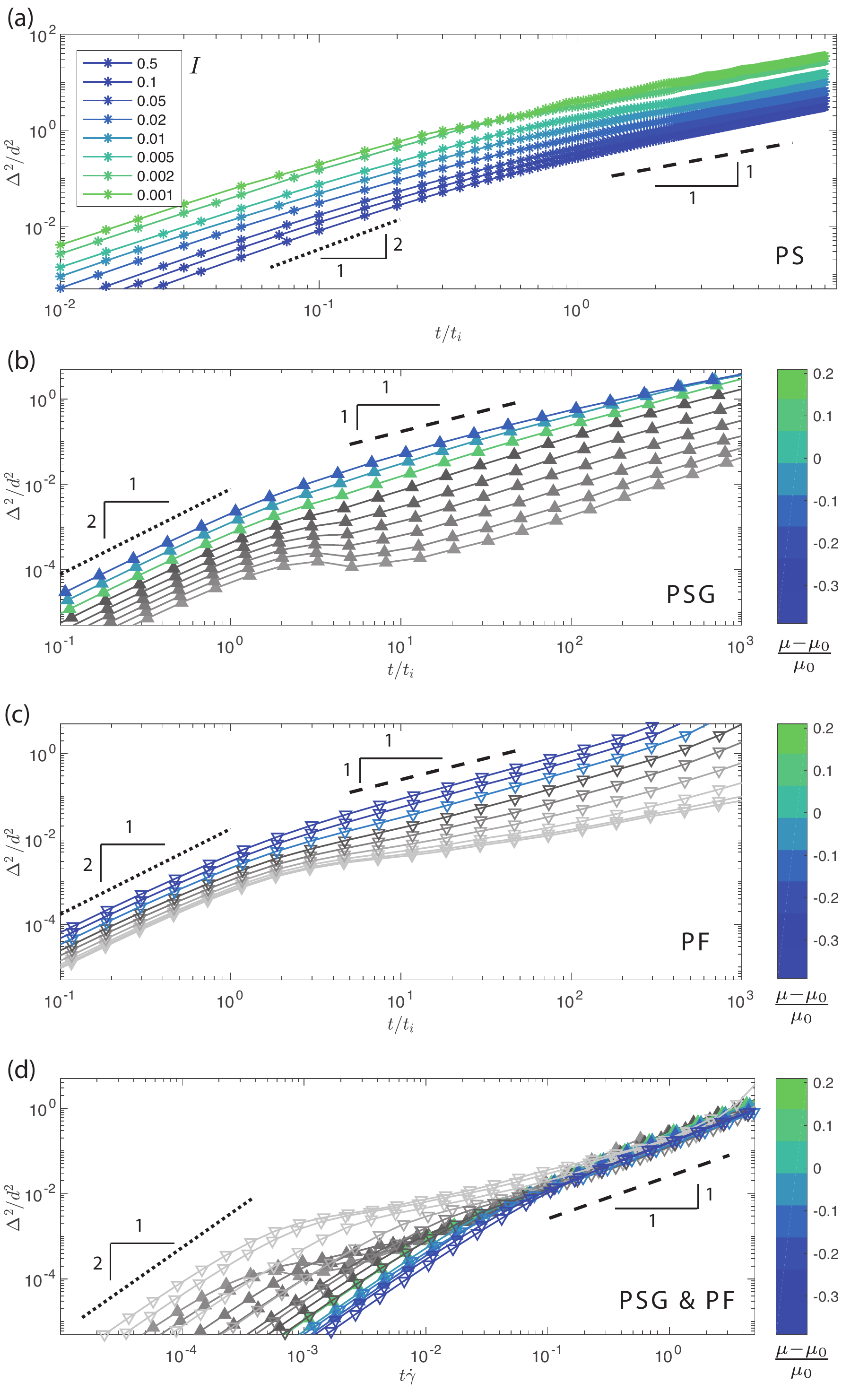}
    \centering
    \caption{Mean squared displacements $\Delta^2(t,y)$ measured at different position $y$ in heterogeneous flows using Eq. (\ref{eq:MSD}). (a) Homogeneous shear flow with $H/d= 120$ at various inertial number from \cite{PhysRevLett.119.178001}, (b) PSG with $H/d=30$, $P_w/E = 400$, $v_w\sqrt{\rho/P_w} = 0.316$ ,  $g/d\sqrt{\rho/P_w} = 0.019$. (c) PF with $H/d=40$, $P_w/E = 1000$, $g/d\sqrt{\rho/P_w} = 0.036$. (d) Combined data from (b) and (c), normalising the time by shear rate time scale $\dot \gamma^{-1}$. In graphs (b-d), the colour scheme represents the level of stress $\mu(y)$ in the layer where the MSD is measured. Green to blue colours represent layers above the yield ($\mu(y)>\mu_0$), while grey shades represent layers below the yield ($\mu(y)<\mu_0$) where the local constitutive law predicts no flow, implying no shear-induce diffusion.  
    } 
    \label{fig:nlocal_msd}
\end{figure}

Plane shear with no gravity presents the advantage of producing homogeneous stresses and shear rate throughout the system. Bi-periodic boundary conditions are used to avoid walls and the shear heterogeneities they induce \cite{rognon2015long}. 
A series of steady flows were performed prescribing a constant normal stress $P=10^{-3}E$ and different values of shear rate $\dot \gamma$. These simulations enabled us to measure the local constitutive law $\mu(I)$ of the materials by averaging the shear to normal stress ratio and the inertial number spatially in the entire flow, and temporally over $25$ shear deformations. The measured values of $\mu$ versus $I$ are shown in \cref{fig:sim_geo} (b). They are best fitted with the local constitutive law (\ref{eq:local}) using $\mu_0 = 0.26$ and $b=1.1$, which is consistent with previously reported values \cite{da2005rheophysics}. It is worth noting that flow occurs only if $\mu>\mu_0$ in this geometry.

Unlike homogeneous plane shear flows, PSG and PF involve some stress gradient that leads to a non-homogeneous shear and a spatial variation of the inertial number in one direction.

PSG is simulated in a periodic domain along $x$ axis, while parallel walls bound the system in the $y$ direction (see Figure \ref{fig:sim_geo}c). Walls are made of  aligned contacting grains with average diameter $2d$, which effectively prevents wall slip \cite{kamrin2012nonlocal}. Wall grains do not rotate and move as a rigid body. The top wall can translate in both directions to prescribe a shear velocity $v_w$. It is also subjected to a vertical external normal stress $P_w$. The vertical wall motion is governed by an inertial dynamics. Its acceleration $\ddot y$ is given at any time during the flow as $ M \ddot y = Ld (P_i-P_w)$ where $L$ is the length of the wall, $M$ the total mass of the wall grains, and $P_i$ the internal normal vertical stress due to contacts between wall and flowing grains. In steady states, the wall vertical position is nearly constant, with some fluctuations smaller than $d$. The bottom wall is immobile. Flowing grains are  subjected to gravity $g$ and the corresponding body force $f_b = \pi d^3 \rho g/4$ in the direction transverse to the shear. This produces a gradient of normal stress in the $y$ direction, while the shear stress is constant. The stress ratio $\mu$ is thus maximum at the top and minimum at the bottom, and it is possible to tune the external normal stress $P_w$ and shear velocity $v_w$ in such a way that the flow is comprised of a top layer that is above the yield  ($\mu(y)>\mu_0$) and a bottom layer that is below the yield ($\mu(y)<\mu_0$). The local constitutive law predicts that there should be no flow in this layer, and therefore no diffusion. However, figure \ref{fig:sim_geo} (d) shows that the inertial number is in fact not null in this layer, owing to non-local effects. This suggests that there may be some diffusion in this layer.

PF is also simulated in a periodic domain in the flow direction and between two parallel walls. Unlike PSG, walls do not move in the flow direction and produce no shear. Both walls are subjected to an external normal stress $P_w$ and are free to move in the $y$ direction according to an inertial dynamic similar to that used in PSG. A body force $f_b$ is applied on flowing grains, but this time in the flow direction $x$. This leads to a gradient of shear stress in the $y$ direction, while the normal stress is constant. As a result, the stress ratio is maximum near the walls and minimum at the centre.  The body force and the applied pressure can be tuned in such a way that a middle layer develops that is below the yield ($\mu(y) <\mu_0$), while the two layers near the walls are above the yield ($\mu(y)>\mu_0$). 
Like PSG, figure  \ref{fig:sim_geo} (f) shows that non-locallity induces some flow in the central zone while it is below the yield, suggesting possible diffusion.

%
%
%

\section{Caging in sub-yield layers}

To highlight the diffusive behaviour in these flows, we measured the typical grain trajectory characterised by their mean square displacement. We took advantage of the time invariance of the steady flow and of their spatial homogeneity (at least in one direction), to measure an averaged mean square displacement defined as:

\be \label{eq:MSD}
 \Delta^2 (t) = \frac{1}{MN}\sum_{i=1}^{N} \sum_{j=1}^{M}\left(y_i(t_j) - y_i(t_j+t)\right)^2,
\ee

\noindent where $t$ is a time interval, $t_j$ a reference time and $y_i$ the $y$ position of grain $i$ at a given time. Averaging is performed considering a series of $M = 100$ reference times selected at random during steady flows. It is also performed on $N$ grains. In homogeneous plane shear flows, all grains can be included in this average, leading to a single MSD for one given flow. In contrast, it is expected that the MSD might depend on the initial position of the grains $y(t_i)$ in heterogeneous flows. MSDs $\Delta^2(t,y)$ are then measured at different position $y$ by averaging on grains located within strips of width $d$ centred at $y$.  

Figure \ref{fig:nlocal_msd} shows examples of MSD evolutions at different layers within flows in the PS, PSG and PF geometries. It appears that these MSDs first exhibit a power law $\Delta^2(y,t) \propto t^{2} $ at short time scales. This denotes  a super-diffusive behaviour, reflecting a ballistic (or constant speed) grain trajectory, as observed in \cite{radjai_turbulentlike_2002,PhysRevLett.119.178001}. In contrast, MSDs exhibit a normal-diffusive behaviour $\Delta^2(y,t) \propto t $ at long time scales. A coefficient of self-diffusion $D$ can be measured in this regime using the Einstein formula \cite{einstein_theory_1906}: 

\be
\Delta^2(y,t) =2D t
\ee

\noindent Figure \ref{fig:nlocal_msd} indicates that this normal-diffusive behaviour develops after a period of time proportional to the shear time $\dot \gamma^{-1}$. Seemingly, normal diffusive behaviour arises after a approximately a tenth of shear deformation ($t\dot\gamma \gtrsim0.1$) in all layers and in all flow geometries. Then, the value of mean square displacement is also similar in all cases,  approximately $\Delta^2\approx 10^{-2} d^2$, which corresponds to a typical grain displacement of $10^{-1}d$. 

Most importantly, MSDs exhibit two qualitatively different behaviour in layers above and below the yield. Above the yield, the super-diffusive regime is directly followed by the normal-diffusive regime. In contrast, below the yield, a sub-diffusive regime develops after the super diffusive phase. This sub-diffusive phase is characterised by a plateauing of the MSD, which denotes a caged trajectory of the grains \cite{marty_subdiffusion_2005,scalliet_cages_2015}. 

This caging dynamics only develops is sub-yield layers. In particular, it does not develop in homogeneous plane shear, even at low inertial numbers. As a consequence, this caging appears to be a distinguishing feature of the grains trajectories in sub-yield layers.

\begin{figure}[]
    \centering
	\includegraphics[width=\columnwidth]{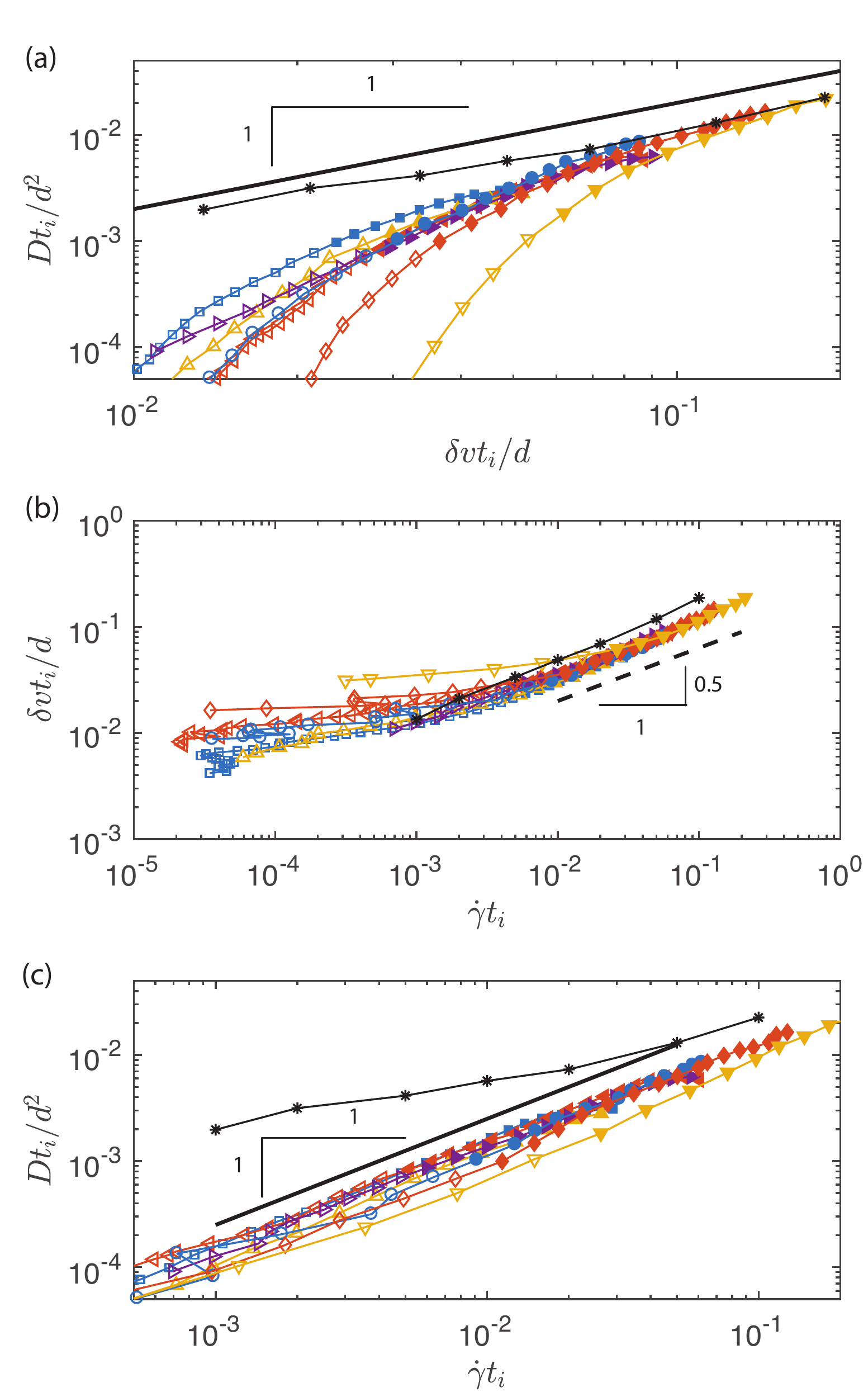}
    \centering
    \caption{Diffusivity and velocity fluctuation scalings in heterogeneous flows. Each symbols represent quantities measured in a given flow at a given position $y$. Symbol code is given in Table \ref{tab:config}. Filled symbols correspond to layers above the yield ($\mu(y)>\mu_0$) and open symbols correspond to sub-yield layers  ($\mu(y)<\mu_0$).
 } 
    \label{fig:diffusion}
\end{figure}
\section{Diffusivity and velocity fluctuations scaling}

As a way to assess the validity of the scalings (\ref{eq:Dscaling}) and (\ref{eq:Vscaling}) in heterogeneous flows, we have measured the profiles of diffusivities $D(y)$ and velocity fluctuations $\delta v(y)$ within  flows in the PSG and PF geometries. Figure \ref{fig:diffusion} shows how these quantities scale with one another, and with the local inertial time $t_i(y)$ and shear rate $\dot\gamma(y)$. These results point out the following three observations.

The first observation is that the scaling (\ref{eq:Dscaling}) of the diffusivity with the velocity fluctuation is not valid everywhere in heterogeneous flows. This is evidenced on figure $\ref{fig:diffusion}a$. This scaling is  valid for layers with the highest velocity fluctuations where results indicate $Dt_i d^2  \approx 0.1 v t_i/d$, which is equivalent to $D \approx 0.1 \delta v d$. These layers correspond to the zone of the flow above the yield. In contrast, there is a neat breakdown of this scaling in sub-yield layers. 

The second observation is that the velocity fluctuation scaling (\ref{eq:Vscaling}) is not valid everywhere in heterogeneous flows. This is evidenced on figure $\ref{fig:diffusion}b$, which suggests two  limits. At high inertial numbers, data seemingly converge toward  the scaling  $\delta v t_i/d \propto I$, or  $\delta v \propto d \dot \gamma$, which is similar to that measured in homogeneous shear flows in this range of inertial numbers. In layers with low inertial numbers, which are sub-yield layers, results suggest that the velocity fluctuations become independent of the shear rate and controlled by the inertial time:
 \be \label{eq:vfluc_NL}
 \delta v \propto d/t_i.
 \ee
 
\noindent This scaling differs from one measured in homogenous shear flow in this range of inertial number. It indicates that velocity fluctuations do not vanish in sub-yield layers even when the shear rate tends to zero. They would vanish in homogeneous plane shear, according to (\ref{eq:Vscaling}).

The third observation is that there is a simple scaling between the diffusivity and the local shear rate in all layers of all tested PSG and PF flows. This scaling, evidenced on figure $\ref{fig:diffusion}c$, is:

\be \label{eq:DNLscaling}
D \approx 0.1 d^2 \dot \gamma
\ee

\noindent It differs from the diffusivity scaling measured in homogenous plane shear flows in this range of inertial numbers, which is $D \propto d^2 \dot \gamma/\sqrt{I}$.

Two conclusions can be drawn from these scalings and from the MSD evolutions. The first conclusion is practical: one can directly deduce the profile of diffusivity in a heterogenous granular flow from the shear rate profile, according to (\ref{eq:DNLscaling}). 

The second conclusion concerns the physical process controlling the diffusivity.  In homogeneous shear, grain velocity fluctuations is controlling their diffusion. The underlying mechanisms is a random walk with a step size proportional to $d$ and a frequency proportional to $\delta v /d$. In sub-yield layers, the diffusivity is controlled by a different mechanisms. Grains still undergo a random walk of step size proportional to $d$, as evidenced by the MSD. However, the elementary step of this walk is comprised of a fast inertial displacement of typical velocity $\delta v = d/t_i$ lasting a period of time proportional to $t_i$, and of a subsequent caging phase. In average, grains are uncaged at a frequency driven by the local shear rate $\dot \gamma(y)$, so that the caging last approximately $\dot \gamma^{-1} - t_i$. As a consequence, the intensity of the velocity fluctuations are no longer influencing the diffusivity in sub-yield layers.

\begin{figure}[!t]
    \centering
	\includegraphics[width=1\columnwidth]{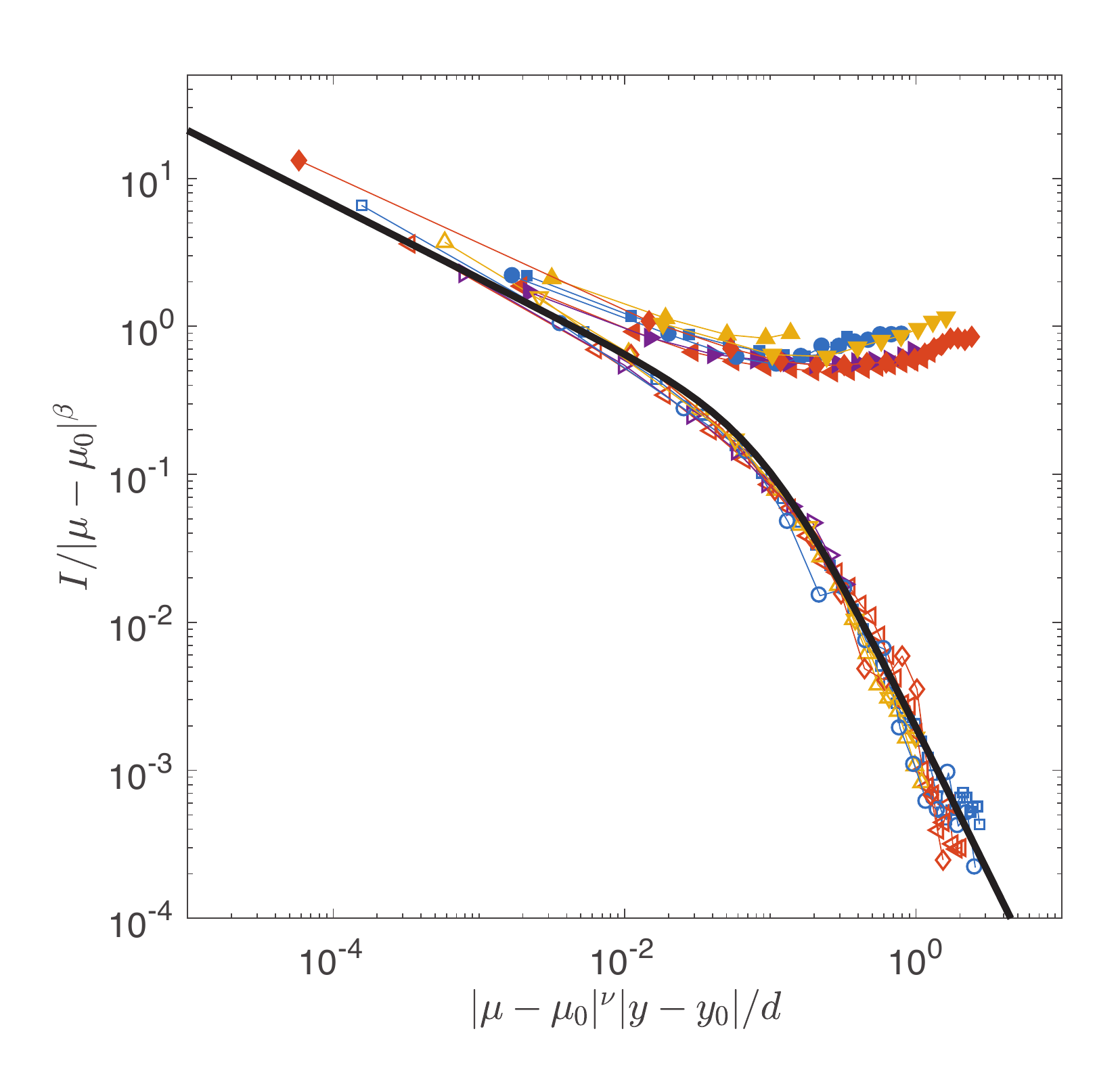}
    \centering
    \caption{Scaling description for dense granular flow. (a) shear stress ratio as a function of inertial number. (b) distance to yield $|y-y_0|$ as a function of inertial number. (c) rescaling of the data from (a) and (b) according to \cref{eq:granular} with $\beta =1$ and $\nu=1$. All symbols represent measurements from DEM simulations with parameters shown in \cref{tab:config}. In (c) the solid curve represents (\ref{eq:fit}) with $A=0.002$ and $B=0.03$.} 
    \label{fig:scaling}
\end{figure}

\begin{figure}[th]
    \centering
	\includegraphics[width=\columnwidth]{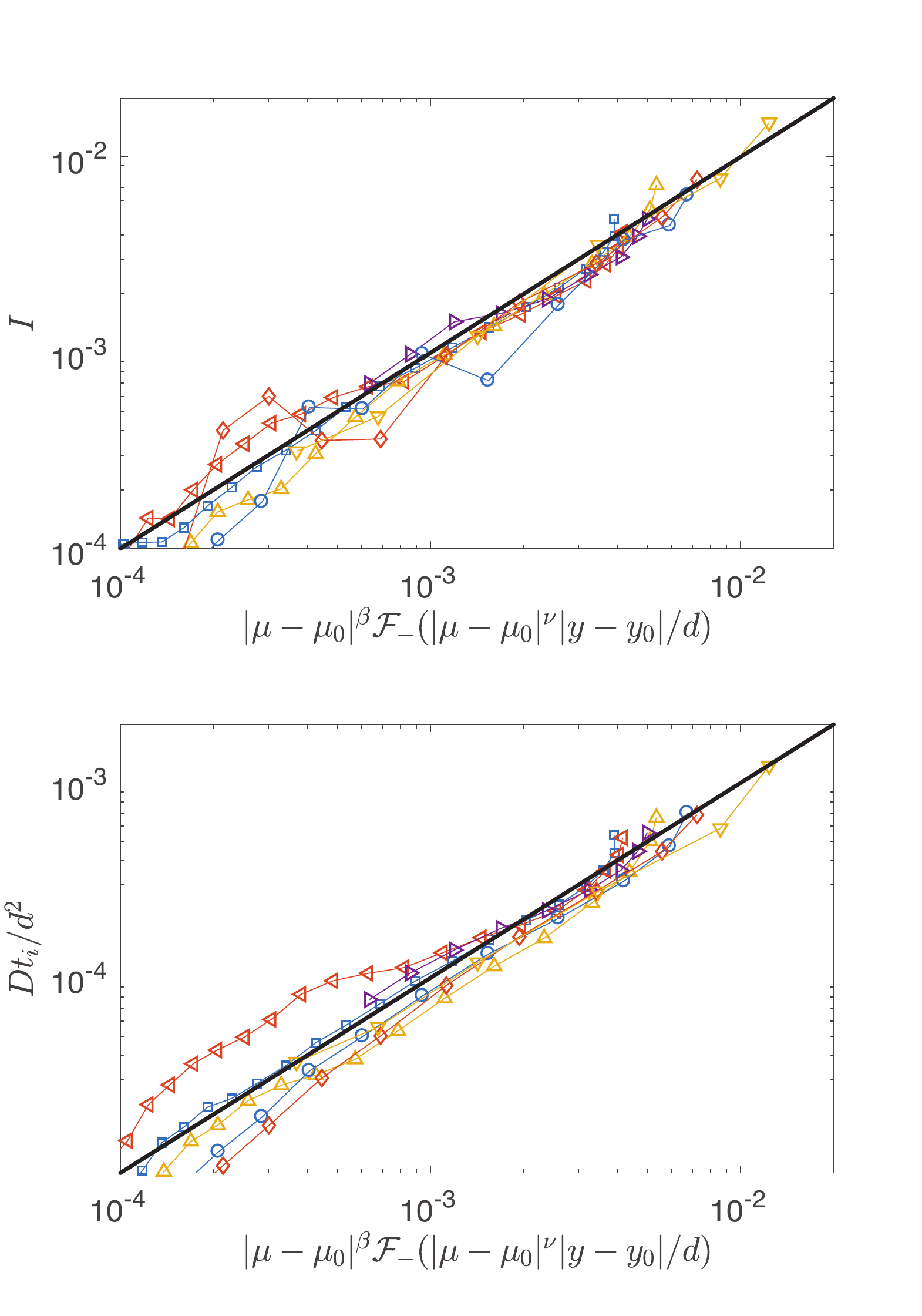}
    \centering
    \caption{Prediction of inertial number and diffusivity in sub-yield layers of PSG and PF flows, using stress condition $|\mu-\mu_0|$ and position $|y-y_0|$. Here, $\beta=1$, $\nu=1$, and coefficients of (\ref{eq:fit}) are $A=0.002$ and $B=0.03$.  $\mathcal{F}_-$  is defined in (\ref{eq:fit}).} 
    \label{fig:prediction}
\end{figure}

\section{Inertial number scaling across the yield} 

We now seek to identify a formula that explicits the profiles of shear rate $\dot \gamma$ that are driving the diffusivity, as per Eq. (\ref{eq:DNLscaling}). The aim is to express  the shear rate profile in terms of local stresses and position in the flow to ultimately infer the diffusivity profiles from these parameters. 

A possible approach to predict the shear rate profiles is to combine the local constitutive law (\ref{eq:local}) with a non-local model \cite{pouliquen_non_local_2009,  kamrin2012nonlocal, bouzid2013nonlocal}. However, existing non-local models are expressed in the form of a PDE and their predictions rely on a choice of boundary conditions, which does not always have a clearly established rationale.

We follow here an alternative approach that has been recently proposed for amorphous solids, referred to as \textit{scaling description} \cite{clark2018critical,Jagla:2017eb,gueudre_scaling_2016}. 
These materials satisfy a Hershel-Buckley local constitutive law: they yield above a shear stress threshold ($\tau>\tau_0$), and then start to flow with a shear rate $\dot\gamma \propto (\tau-\tau_0)^\beta$, where $\tau_0$ and $\beta$ are material dependent parameters. Interestingly, amorphous solids also exhibit non-local effects that are similar to those in granular flows: some flow may exist in a zone below the yield near if an adjacent zone that is flowing. The scaling description of such non-local effects consists in establishing a scaling for the local shear rate in terms of the distance to the yield, as follows \cite{gueudre_scaling_2016,clark2018critical}:
\begin{equation}
\dot\gamma (y) = |\tau-\tau_0|^\beta \mathcal{F}_{\pm}(|\tau-\tau_0|^\nu|y-y_0| )
\label{eq:amorphous}
\end{equation}
\noindent where $y$ is the position of layer, $y_0$ is the position of the layer in the flow where $\tau(y_0)=\tau_0$, and $\nu$ is some exponent. $\mathcal{F}_+$ and $\mathcal{F}_-$ are referred to as scaling functions, which correspond to layers above ($\tau>\tau_0$) and below ($\tau<\tau_0$) the yield, respectively. 

We seek to adapt here this approach to granular flows by considering the relevant frictional yield criteria and non-dimensional shear rate $I$, as:

\begin{equation}\label{eq:granular}
I (y) = |\mu-\mu_0|^\beta \mathcal{F}_{\pm}(|\mu-\mu_0|^\nu|y-y_0| /d)
\end{equation}

\noindent  Figure \ref{fig:scaling} shows that this scaling does capture the measurements in our simulated PSG and PF granular flows using $\beta = 1$ and $\nu=1$:  data collapse on two curves, one for layers below the yield and one for the layers above the yield. Above the yield, $\mathcal{F}_-+$ becomes seemingly constant and of the order of $1$, indicating that non-local effects may be neglected in these layers. 
Below the yield, data suggests a transition from a power $-0.5$ to a power $-2$ for the function $\mathcal{F}_-(x)$. We introduce the following interpolation to capture these two regimes and their transition:

\bee
\mathcal{F}_- (x) &=& \frac{A }{x^2+B\sqrt{x}} \label{eq:fit}\\
x &=& |\mu-\mu_0| |y-y_0|/d
\eee
\noindent where $A$ and $B$ are the two fitting parameters. \cref{fig:scaling} shows that this function satisfactorily captures the measurements with $A=2\times10^{-3}$ and $B=0.03$.
Interestingly, $x$ may be seen as a distance to the yield that includes a stress-wise distance $|\mu-\mu_0 |$ and an Euclidian distance $|y-y_0|/d$. 
Far from the yield, for $x\gg1$, the function $\mathcal{F}_- (x)$ tends to $\mathcal{F}_- (x) = Ax^{-2}$. Considering (\ref{eq:granular}), the profile of inertial number is then given by:

\be \label{eq:I_scaling}
I (y) \approx  \frac{A}{|\mu-\mu_0|} \frac{d^2} {|y-y_0|^2}
\ee 

\noindent Figure \ref{fig:prediction} shows how the profile of inertial number and the profile of diffusivity can be captured using the scaling prediction for the inertial number (\ref{eq:granular}) and (\ref{eq:DNLscaling}), in all layers of the heterogeneous flows. 

We note that the scaling  (\ref{eq:I_scaling}) is consistent with the \textit{self-activated} mechanism underlying the non-local model introduced in \cite{pouliquen_non_local_2009}. This mechanism considers that plastic events may be triggered in sub-yield layers by stress fluctuations originating from remote flowing layers. In this model, it is though that stress fluctuation decays as distance to the power $-2$ from their origin. This is consistent with the scaling $I (y) \propto  \frac{d^2} {|y-y_0|^2}$. Further still, it is though that the magnitude of theses stress fluctuations required to uncaged a grain is proportional to $|\mu-\mu_0|$, which is consistent with the scaling $I (y) \propto  \frac{1}{|\mu-\mu_0|}$.

\section{Conclusions}

This study points out that shear-induced diffusion is qualitatively different in granular layers flowing below and above the yield.

Above the yield, diffusivity is proportional to the velocity fluctuations, which themselves are a driven by the shear rate and possibly by the inertial number, as per (\ref{eq:Dscaling}) and (\ref{eq:Dabove}).
In contrast, diffusivity in sub-yield layers is not controlled by velocity fluctuations, and velocity fluctuations become shear-rate independent and controlled by the inertial time. 

Our results indicate that, below the yield, diffusivity is directly proportional to the local shear rate, as per (\ref{eq:DNLscaling}). This shear rate profile may be deduced from non-local continuum models, or by the scaling approach we introduced, which led to (\ref{eq:I_scaling}). These scalings can readily be used to resolve diffusion processes in non-local granular flows. 

Our results also point out the emergence of a caging dynamics in sub-yield layers. This suggests that such a caging dynamics could be used as an indicator of whether a sheared layer is below or above the yield. This could be used to measure the yield stress $\mu_0$ directly from non-local flows when this quantity is not known a priori.

\end{document}